# Real Time Quantum Dynamics of Spontaneous Translational Symmetry Breakage in the Early Stage of Photo-induced Structural Phase Transitions

**Keiichiro Nasu**

Institute of Material Structure Science, KEK, 1-1, Oho, Tsukuba, 305-0801, Japan; knasu@post.kek.jp
* Correspondence: keiin@hotmail.co.jp; Tel.: +81-297-72-7547

## Abstract:

Real time quantum dynamics of the spontaneous translational symmetry breakage in the early stage of photoinduced structural phase transitions is reviewed and supplementally explained, under the guide of the Toyozawa theory, which is exactly in compliance with the conservation laws of the total momentum and energy. At the Franck Condon state, an electronic excitation just created by a visible light, is in a plane wave state, extended all over the crystal. While, after the lattice relaxation having been completed, it is localized around a certain lattice site of the crystal, as a new excitation. Is there a sudden shrinkage of the excitation wave function, in between. The wave function never shrinks, but only the spatial, or inter lattice site quantum coherence, interference of the excitation disappears, as the lattice relaxation proceeds. This is nothing but the spontaneous breakage of translational symmetry.

---

## 1. Introduction

The spontaneous symmetry breakage is one of the most important problems of great interests in the solid state physics for these several decades. As already well-known, this problem is closely related, not only to the various mechanisms of crystalline magnets, but also to the BCS mechanism of the superconductivity, and even to the Higgs mechanism of the elementary particle physics [1].

The mechanism for the ferromagnetism of itinerant electrons in a conductive crystal within the mean field approximation [2], is most easy for us to understand the spontaneous symmetry breakage. At first, we start from a hypothetical paramagnetic state of itinerant electrons. It is perfectly symmetric, in the sense that un-spin electrons and down-spin ones equally occupy all the lattice sites of the crystal, resulting in no macroscopic magnetic (spin) moment, without an externally applied magnetic field. In the next, we hypothetically assume a spatially uniform but finite unequal occupation. Under this condition, we estimate the total free energy of the system within the mean field theory. Finally, we determine the real value of this hypothetical finite unequal occupation, so that it will give the lowest free energy. If this lowest energy is even lower than the starting paramagnetic state, without an externally applied magnetic field, we can get a ferromagnetic state which has a spontaneous and macroscopic magnetic (spin) moment. Thus, we can get the symmetry breaking in the space of the electron spin.

It should be noted that, during this symmetry breaking transition from the paramagnetic state to the ferromagnetic one, the whole system is assumed to be always in the thermal equilibrium, and hence, the speed of the transition has to be infinitely slow, according to the principle of the thermodynamics.

Keeping this point in mind, let us now proceed to the optical region spectroscopy of insulating crystalline solids. In this research field, according to the rapid progress of time resolved laser techniques, real time quantum dynamics of optically created electronic excitations is gradually clarified in detail up to a pico- or femto-second time scale. This advantageous experimental technology has also been intensively applied even to the present spontaneous symmetry breaking problem. As a result, experimental and theoretical studies for this problem have been intensively developed, although it is quite different way than mentioned above. That is, the real time quantum dynamics of the symmetry breakage.

One of its typical results is the spontaneous ( self- ) localization of an exciton in insulating crystals. The exciton is already well known to be the most elementary optical excitation across the energy gap of insulating crystalline solids [3, 4]. Just after the optical excitation, the exciton is always in a plane-wave state extending all over the crystal. After the lattice relaxation having been completed, however, it is in a localized state, being trapped by the self-induced local lattice distortion around it, provided that the exciton-phonon coupling is short ranged and sufficiently strong. This concept was initiated by Rashba [5] and Toyozawa [6] independently, and also developed afterwards rather independently [4, 7].

This localization is intrinsic in the sense that it occurs without extrinsic trapping potentials, say, due to impurities in the crystal [7]. Thus, it is nothing but the spontaneous translational symmetry breakage. Usually, this self-localized exciton still remains within the energy gap of the original insulating crystal, and is luminescent. Hence, it finally disappears after radiating another photon whose energy is a little smaller than that used for the initial excitation [8]. However, if the exciton-phonon coupling is further strong, it remains frozen as a non-luminescent localized electronic excited state with a large lattice distortion round it [9].

One can now say, it is a tiny photo-induced structural phase transition (PISPT). As already well known, there discovered a new class of many solids, which, being shone only by visible photons, become pregnant with a macroscopic excited domain that has new structural and electronic orders quite different from the starting ground state [10, 11]. This phenomenon is called PISPT [10], and the present frozen non-luminescent localized electronic excited state is nothing but a PISPT, although the domain size of the new phase is the possible minimum.

The purpose of the present paper is to review and supplementally explain this spontaneous translational symmetry breakage in the very early stage of the PISPT. It was once reviewed only shortly [12], and the explanation was also quite insufficient.

## 2. Adiabatic Nature of Exciton Self-localization

As shown by Toyozawa [9], the PISPT phenomenon is closely related to the aforementioned self-localization of an exciton in an insulating crystal. It can be simply described by the following model Hamiltoninan ( $\equiv \mathrm{H_F}, \hbar = 1$) for an exciton,

$$\mathrm{H_F} = -\mathrm{T_F} \sum_{<l,l^*(\neq l)>} [\, \mathrm{F}_{l^*}^{+}\mathrm{F}_l + \mathrm{h.c.}\,] + \sum_l \left(\mathrm{E_g} + 6\mathrm{T_F} - \omega_0 \mathrm{S Q}_l\right)\mathrm{F}_l^{+}\mathrm{F}_l + \frac{\omega_0}{2}\sum_l \left(-\frac{\partial^2}{\partial \mathrm{Q}_l^2} + \mathrm{Q}_l^2\right). \quad (1)$$

Here, $T_F$ ( > 0 ) is the resonant transfer (energy) of an exciton from a lattice site $l$ to its nearest neighbouring sites $l^*$ in a simple cubic crystal. The bracket $< l, l^* >$ in Equation (1) denotes that these two lattice sites $l$ and $l^*$ are nearest neighbours with each other. $F_l^+$ in Equation (1) is the creation operator of this exciton at the lattice site $l$. It is not the charge transfer type excitation, but a Frenkel type ( intra-atomic, or intra-molecular ) one well localized only in each lattice site. As schematically shown in Figure 1(a), $E_g$ in Equation (1) denotes the energy gap of this insulator, while S is the dimensionless coupling constant of this exciton to a site localized phonon, of which energy and dimensionless coordinate are $\omega_0$ and $Q_l$, respectively. In this section, the kinetic energy of this phonon is negelected, because of the adiabatic approximation. Usually, $E_g$, $6T_F$ and $\omega_0 S$ are quantities of the order of eV, while $\omega_0$ is 10meV or so.

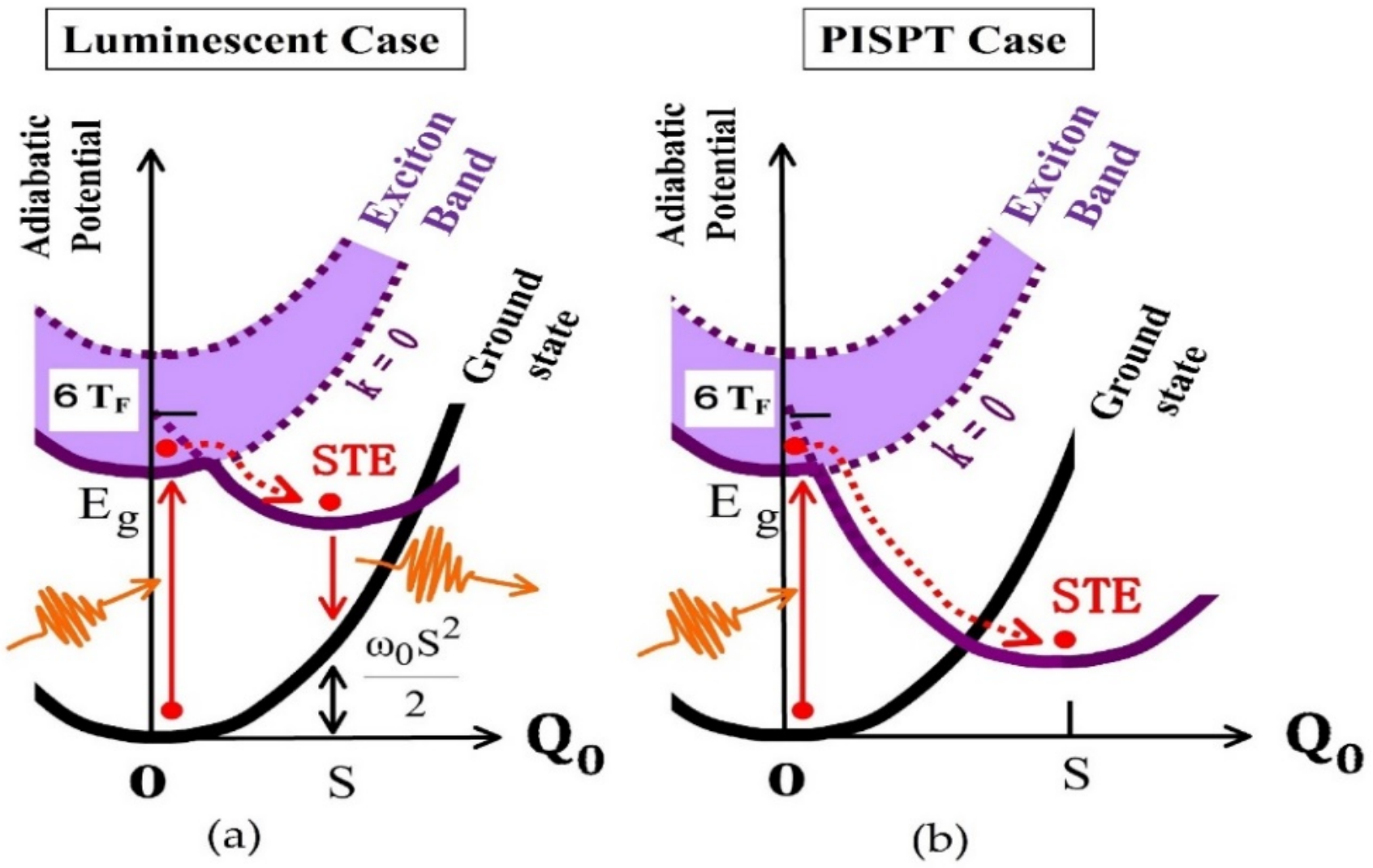

**Figure 1.** The adiabatic potential energy surface of an exciton, at the Franck-Condon excited state ( the red upward vertical arrows ), and the STE, as a function of the $Q_0$. (**a**) The Luminescent case. (**b**) The PISPT case.

Within the adibatic approximation, the eigen-state ( $\equiv |\Psi(Q_l)>, <\Psi|\Psi> = 1$ ) of this $H_F$ will be given as a function of $Q_l$. It is unknown at present, but we determine it under the condition that the total number of the exciton is just one, $\sum_l F_l^+ F_l = 1$. After formally taking the average of $H_F$ with respect to this unknown $|\Psi>$, we can apply the Hellmann-Feynman theorem to Equation (1), and can get as,

$$\frac{\partial <\Psi|H_F|\Psi>}{\partial Q_l} = 0, \qquad <\Psi|F_l^+ F_l|\Psi> = \frac{Q_l}{S}. \tag{2}$$

Substituting this Equation (2) into the original Equation (1) , we also get

$$< H_F >= (E_g + 6T_F) - T_F \sum_{<l,l^*(\neq l)>} [< F_l^+ F_{l^*} > + < F_{l^*}^+ F_l >] - \frac{\omega_0 S^2}{2} \sum_l < F_l^+ F_l >^2 \quad (3)$$

where $\Psi$ is omitted in the averages $< \cdots >$, for simplicity. We should note that this Equation (3) holds only at local minimum ( or extremum ) points in the multi-dimentional coordinate space spanned by $Q_l$, since it is obtained by using Equation (2).

When the exciton-phonon coupling is sufficiently strong, $6T_F < (\omega_0 S^2)/2$, according to Shinozuka and Toyozawa [7], we have only two types of minima in the adiabatic potential energy surface of the excited state, as schematically shown in Figure 1(a). The first minimum is the globale one with $< F_l^+ F_l > = \delta_{l,0}$, being localized, say, at the origin $l = 0$ with a large lattice displacment, $Q_0 = S$. Its electronic energy $\left((E_g + 6T_F) - \omega_0 S^2\right)$, given by the second term of Equation (1), formally starts from the exciton band center $(E_g + 6T_F)$, but goes below the exciton band, as a local lattice dispacement $Q_0$ is self-induced ( $0 \to S$). It is called self-trapped（, or self-localized ）exciton (STE) state.

The second local minimum is $< F_l^+ F_l >= 1/N$, where N denotes the total number of the lattice sites in the crystal. This is the plane-wave state of the exciton whose wave-vector ($\equiv k$) is zero, k = 0, and its energy is just the energy gap $E_g$. Thus, the final state of the Franck-Condon (FC) excitation by light ( the red vertical upward arrows ), is this plane wave state, being the lowest one within the exciton band, as shown in Figure 1. While, after the lattice relaxation, as schematically shown by the dashed red allows in Figure 1, the whole system reach the STE state. We should also note that, at this largely displaced lattice configuration, even the elastic energy of the ground state, as well as that of the STE, increases upto $\omega_0 S^2/2$, since the lattice distortion (, the last term of Equation (1), ) is common to all states. If the total energy of this STE state is above the ground state one at this lattice configuration,

$$(E_g + 6T_F) - \frac{\omega_0 S^2}{2} > \frac{\omega_0 S^2}{2}, \quad (4)$$

this STE state still remain in the gap of this insulating crystal, and finaly disappears with a luminescence, of which energy is a little smaller that the exciting one, as shown in Figure 1(a). This is the ordinary situation widely realized in luminescent insulators [8].

As shown in Figure 1(b), however, if the exciton-phonon coupling is so large as to relax down even lower than the ground state at this largely displaced lattice configuration,

$$(E_g + 6T_F) - \frac{\omega_0 S^2}{2} < \frac{\omega_0 S^2}{2} \quad (5)$$

the system becomes non-luminscent, and the STE remains forever within the adiabatic approximation at absolute zero temperature. This is nothing but the start of the PISPT [10], although the domain size of the new phase is the possible minimum.

Thus, we have seen the spontaneous translational symmetry beakage. Similar to the above Stoner theory [2], its mechanism is also a sufficient energy lowering from the perfectly symmetric state.

According to the adiabatic principle, however, the speed of this symmetry breaking transition is also infinitely slow.

Incidentally, within the framework of the present theory, we can formally encouter an extremely strong coupling case that the enegry of the STE becomes even lower than the starting ground state itself; $\left(E_g + 6T_F - \omega_0 S^2/2\right) < 0$. We can not use Equation (1) for a such contradicting case.

**3. Dynamics of Self-localization**

Let us now proceed to the non-adiabatic quantum dynamics of self-localization, including the kinetic energy term of the phonon in Equation (1). The wavelength of visible light is quite longer than the lattice constant of the crystal. This means that the wave vector of the visible photon is almost zero, because it is extremely smaller than the other wave vectors of an exciton in the first Brillouin zone of this crystal. Consequently, as already mentioned in the previous section, the initial FC type excited state ($\equiv |FC>$) is the Bloch wave whose total wave vector ( $\equiv k$ ) is almost zero, having the same translational symmetry as that of the original crystal. It is given by

$$|FC> = N^{-\frac{1}{2}} \sum_l e^{-ik\cdot l} F_l^{\dagger} |0>, \; k \to 0, \qquad |0> \equiv \text{Exciton・phonon true vacuum.} \qquad (6)$$

Thus, the probability density of the exciton at each lattice site of the crystal is inversely proportional to N (volume of the crystal),

$$< FC|F_l^{\dagger} F_l |FC> = 1/N \quad (\text{, the unit of length is the lattice contant }). \qquad (7)$$

Meanwhile, the self-localization mentioned above, is often misunderstood to be a sudden shrinkage of the excitation energy or the excitation wave function from the infinitely extended Bloch state $|FC>$ to a localized one within a lattice site, say, only at the central lattice site of the crystal. This picture of sudden shrinkage, however, is completely wrong. Even if it will shrink, it will do so, not only to the central site, but also to all other sites simultaneously and equally, with a certain transient quantum coherence among them. This is not the shrinkage, any more.

Before, during and even after the self-localization, the wave function never shrinks, as shown by Cho and Toyozawa [13]. They have proposed the following simple but Bloch type self-localized state ($\equiv |STE>$),

$$|STE> = N^{-\frac{1}{2}} \sum_l e^{-ik\cdot l - S(F_l^{\dagger} F_l)\frac{\partial}{\partial Q_l}} F_l^{\dagger} |0>, \quad k \to 0. \qquad (8)$$

In this Bloch type STE state, the density at each lattice site of the crystal is unchanged from Equation (7), and is still inversely proportional to N,

$$< \mathrm{STE}|F_l^{+}F_l\,|\mathrm{STE}> = 1/N\,. \quad (9)$$

As described in Equation (8), however, through the following displacement operator for phonons,

$$e^{-S(F_l^{+}F_l)\frac{\partial}{\partial Q_l}}\,, \quad (10)$$

the self-localized state at each lattice induces a large ( $S \gg 1$ ) lattice distortion only in its site. This phonon displacement will appear and disappear according to the presence or the absence of exciton, since it is just proportional to $F_l^{+}F_l$ . In other words, once this large local lattice distortion occurs, the exciton has heavily dressed in phonons. Hence, even if it tries to move only to a neighboring lattice site from its original one, it has to annihilate all these phonons (larger distortion) and has to make them again at the neighboring site, newly. This phonon dressing picture was also developed by Rashba and his co-workers [14].

The aforementioned limited probability of the spatial motion can be estimated by the inter lattice-site coherence $(\equiv C(\Delta), \Delta \neq 0)$ of exciton, which is given as,

$$C(\Delta) = \sum_l < \mathrm{STE}|\, F_{l+\Delta}^{+}F_l\,|\mathrm{STE}>. \quad (11)$$

It becomes almost zero when the exciton–photon coupling is very strong

$$C(\Delta)\ \rightarrow\ 0\ (,\ =< 0|e^{-S\frac{\partial}{\partial Q_{l+\Delta}}}|0>< 0|e^{-S\frac{\partial}{\partial Q_l}}|0>,\ \ S \gg 1\,)\,. \quad (12)$$

While, at the initial FC state, this inter lattice-site coherence $(\equiv C_{FC}(\Delta), \Delta \neq 0)$ is given as

$$C_{FC}(\Delta) \equiv \sum_l < FC|F_{l+\Delta}^{+}F_l|FC> = 1, \quad (13)$$

and remains finite. Thus, we can say, the spatial, or the inter-site quantum coherence of exciton becomes zero when the exciton–phonon coupling is very strong, although it was finite at the FC state, as schematically shown in Figure 2. This is nothing but the spontaneous translational symmetry breaking, and finally makes a classical and local picture for exciton valid. This relaxation with the large lattice distortion from the Bloch wave to the self-localized one can occur even at absolute zero temperature.

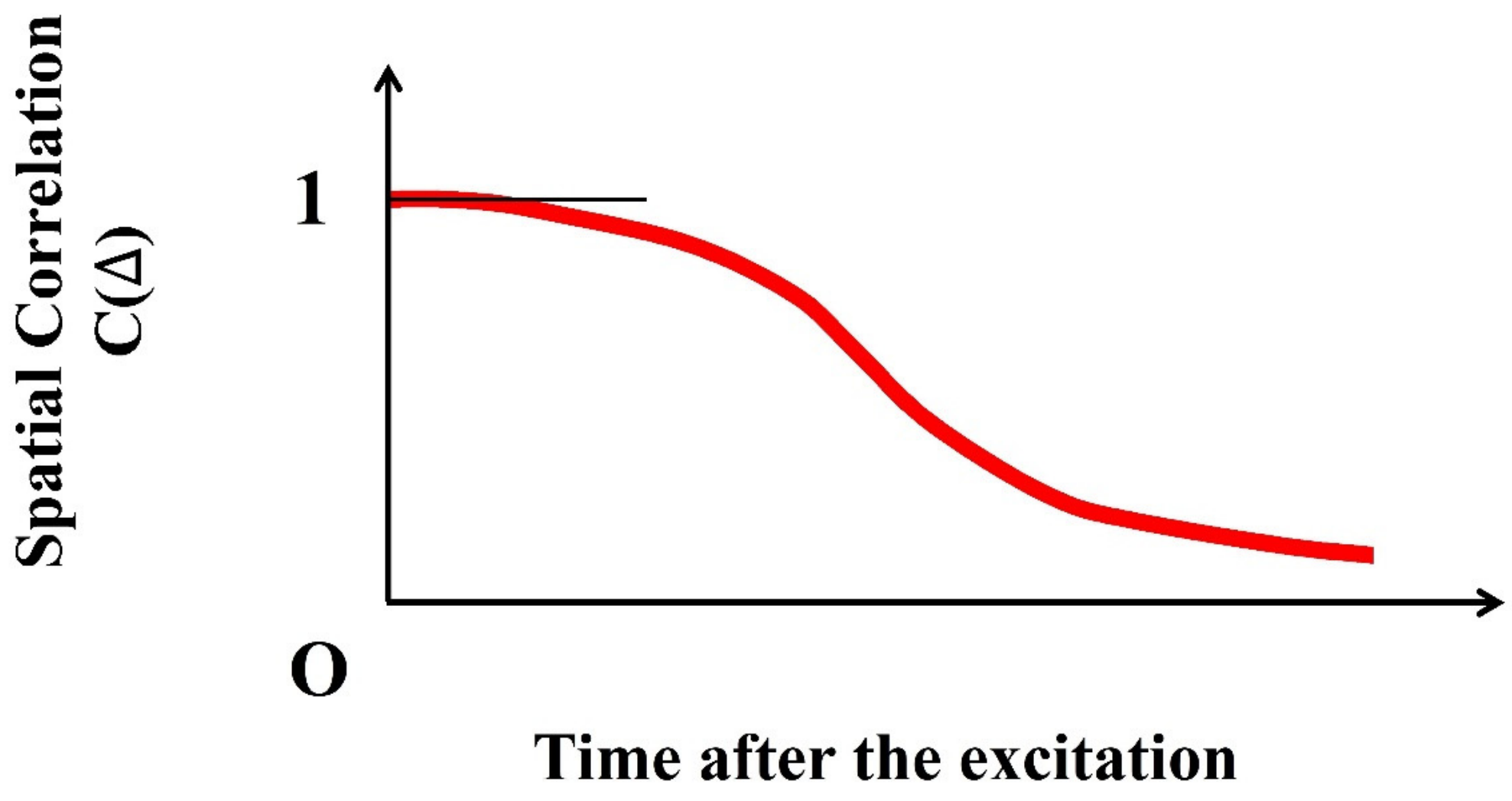

**Figure 2.** The schematic nature of the decrease of spatial correlation of an exciton $[C(\Delta), \Delta \neq 0]$, as a function of the time, after the optical excitation. Nasu and Toyozawa [15] have calculated the rate of this decreasing in detail, using more realistic models.

The above arguments related with Figure 2 for the exciton self-localization, however, are quite formal and too conceptual. For this reason, Nasu and Toyozawa [15] have calculated the rate of this symmetry breaking transition in detail, using more realistic models. This transition is described only in a one-dimensional space spanned by $Q_0$ in Figure 1. However, in reality, it will occur in a multi-dimensional space spanned by many phonon mode coordinates. More-over, the FC sate and the STE are not completely orthogonal with each other. These points are taken into account in the context of the multi-phonon non-radiative transition, and the rate is obtained as a function of the exciton band width, coupling constants of the optical and acoustic phonons, and the exciton Wannier radius. It is in the region from $10^{-1}\omega_1$ to $10^{-2}\omega_1$(, $\omega_1 \equiv$ the averaged acoustic phonon energy), being more probable than the ordinary radiative decay rate of an exciton, in good agreements with the experimental results in alkali iodides and rare gas solids.

As for the PISPT, the self-localization is not the finall distination, but the exciton further proliferates to result in a localized semi-macroscopic domain of a new phase [10]. However, we have to surely pass this early stage dynamics with the spontaneous translational symmetry breakage.

## 4. Conservation Laws of the Total Momentum and Energy, Heat Reservoir, Classical Localization

This early stage dynamics is a purely quantum mechanical one, and hence, it has to be in compliance with the conservation laws of the total momentum and the total energy, exactly. As for the total momentum, being zero from the beginning, can be easily seen to be conserved from Equation (8). While, to see the total energy conservation in detail, we have to tacitly assume a direct transfer type interaction between neighboring site-localized phonons with an interaction constant ( $\equiv \Delta\omega_0,\ 0 < \Delta\omega_0 < \omega_0$ ). Hence, the last term of Equation (1) becomes as,

$$\sum_l \left(-\frac{\partial^2}{\partial Q_l^2} + Q_l^2\right) \to \omega_0 \sum_l \left(B_l^{+} B_l + \frac{1}{2}\right) + \Delta\omega_0 \sum_{<l,l^*\ (\neq l)>} (B_{l^*}^{+} B_l + \mathrm{h.c.}),\ B_l \equiv 2^{-1/2}\left(\frac{\partial}{\partial Q_l} + Q_l\right), \quad (14)$$

although it was not written explicitly in the stage of the section 1.

By this interaction, the energy difference between the FC state and the STE ( shown in Figure 1) is finally radiated as a sound ( or heat) wave. It is schematically denoted by a thin brown circle in Figure 3.

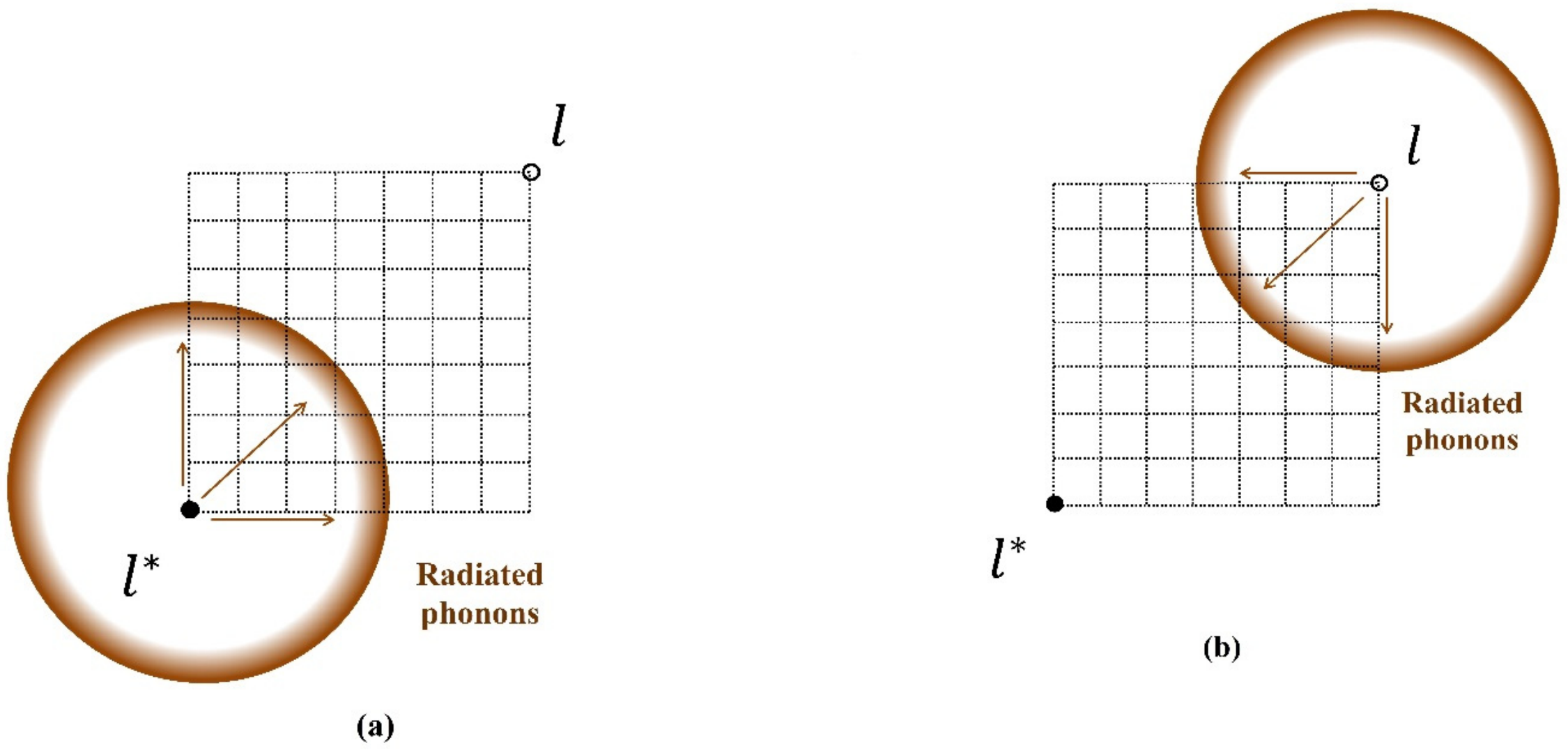

**Figure 3.** The schematic nature of the phonon radiation in the relative space of each STE, at a long time ($\gg \omega_0^{-1}$ ) and a largely distant ( $|l - l^*| \gg 1$ ) limits. The radiated phonon wave front becomes very diffuse and almost spherical around the STE site. (**a**) When the STE is at $l^*$, it is from $l^*$ to many very distant $l$ s. (**b**) Vice verse. The situation of the phonon radiation in the short time ($\sim \omega_0^{-1}$) and shortly distant ( $|l - l^*| \sim 1$ ) region was described by Nasu and Toyozawa [15] in detail, using a more realistic microscopic model for the exciton and the phonons, as well as the couplings among them.

Without this radiation, the symmetry breaking can never be completed. The most important point is that, this radiation of phonons occurs from each STE site to infinitely distant numberless lattice sites, simultaneously and equally. If the STE is at $l^*$, infinitely distant numberless lattice sites are such $l$ s as shown in Figure 3(a), and vice verse, as shown in Figure 3(b). That is, the radiation occurs only in the relative ( or internal ) lattice space, whose central lattice site is occupied by this STE. This is an irreversible process, since this relative space is also infinitely large. While these relative spaces are orthogonal with each other, since

$$< 0|\, F_{l^*} F_l^{+} |0 > = 0, \qquad l^* \neq l. \tag{15}$$

Consequently, a superposition of states realized in each relative space can be possible, just like Equation (8), even though we have included this irreversible phonon radiation. Since this Equation (14) does not change the total energy, but only makes phonons to move, we can now see the total energy, as well as the total momentum, are well conserved.

We can now think of the usual master equation method to describe the lattice relaxation [16]. By this method, however, from the beginning, the whole system is clearly divided into two; a relevant system on which we focus, and a heat reservoir which instantaneously absorbs energies released from the relevant system. By tracing out the reservoir variables, we can thus describe the relaxation dynamics of the relevant system. In the electron-phonon coupled systems, the electronic part is often regarded to be the system, while the phonon is regarded to be the reservoir. As we can easily infer from Figure 3, however, such a priori division is impossible in the present problem. The phonons at infinitely distant lattice sites from the STE may be the heat reservoir, but the central SET site as well as these distant sites are all in the relative space, being not fixed in the real lattice at all.

Incidentally, long after this quantum and spontaneous localization, thus, having been completed, an ordinary classical localization may also occur, since the localized exciton can also slowly and diffusively move and will be trapped at dislocation or rare impurity sites, which unavoidably exist in the ubiquitous crystal.

## 5 Conclusions

Real time quantum dynamics of the spontaneous translational symmetry breakage due to a light excitation in the early stage of photo-induced structural phase transitions is reviewed and explained, under the guide of the Toyozawa theory. At the FC state, an electronic excitation just created by a visible light is in a plane wave state, extended all over the crystal. While, after the lattice relaxation having been completed, it is localized around a certain lattice site of the crystal, as a new excitation. Is there a sudden shrinkage of the excitation wave function, in between? No! The wave function never shrinks, but only the spatial (or inter lattice-site) quantum coherence of the excitation disappears, as the lattice relaxation proceeds. This is nothing but the spontaneous breakage of translational symmetry. We have also reviewed, the roles of the conservation laws of the total momentum and energy, as well as the specific nature of the heat reservoir. A possibility of the final classical localization was also discussed, in comparison with the present quantum and spontaneous localization.